\documentstyle[aps,epsfig,multicol]{revtex}
\begin{document}
\draft
\title{Screening of electrostatic potential in a composite fermion system}
\author{D.V.Fil}
\address{Institute for Single Crystals National Academy of Sciences of
Ukraine,
Lenin av. 60 Kharkov 310001 Ukraine\\
e-mail: fil@isc.kharkov.ua}
\date{December 11, 1998}
\maketitle
\begin{abstract}
Screening of the electric field of a test charge by  monolayer and
double-layer composite fermion systems is considered.
It is shown that the electric field of the test charge
is partly screened at distances
much  larger than the
magnetic length.
The value of screening as a function of the distance depends
considerably on the filling factor.  The effect of variation of the value
of screening in the double-layer system upon a transition to a state 
described by the Halperin wave function is determined.
\end{abstract}
\begin{multicols}{2}
The model of composite fermions was proposed by Jain\cite{1} to describe
the fractional filling factor hierarchy observed in
quantum Hall systems. It was shown in Ref.~\cite{1} that
the Laughlin wave function has a topological structure equivalent to that
of a system of quasiparticles carrying the statistical charge and the
statistical gauge field flux. In the mean field approximation, the
interaction of composite quasiparticles with the gauge field is reduced
to the action of a self-consistent field which partially screens the
external magnetic field. Consequently, the fractional quantum Hall effect
in the electron system emerges as an integer quantum Hall effect
in a system of composite fermions.

Lopez and Fradkin\cite{2} and Halperin, Lee, and Read\cite{3} developed
the Chern--Simons formalism for describing a system of composite fermions.
The formalism allows us to introduce systematically the corrections to
the mean-field solution by expanding the effective Lagrangian in small
deviations of the gauge field from the mean-field configuration.

The approach developed in Refs.~\cite{2,3} was generalized by
Lopez and Fradkin\cite{4} to the case of a double-layer system. A
specific feature of such a system is the possibility of formation of
generalized Laughlin states whose multiparticle wave function is
characterized by an additional set of zeros for coinciding $x, y$
coordinates of electrons in the opposite layers.
The wave function for such states was
proposed by Halperin.\cite{5} Although the original analysis presented
in Ref.~\cite{5} concerns a monolayer system of unpolarized
electrons, the wave function proposed by Halperin is
generalized to the case of a two-layered system, if pseudospins,
which corresponds to the layer index are introduced.
The  states \cite{5} may appear for
new (different from monolayer) filling factors.
In particular, the quantum Hall effect emerges for a filling
factor $\nu = 1/2$, which is indeed observed in experiments \cite{6,7}.
Moreover, for certain fixed values of the filling  factor in two-layered
systems, a phase transition between different generalized  Laughlin states
becomes possible upon a change in the separation between the layers (see,
for instance,\cite{8})

In this work, we study the effect of screening of the external electrostatic
potential by a composite fermion system.
We consider the screening of the field of a test charge
located at the boundary of a semi-infinite medium with the dielectric
constant $\epsilon$, having at a certain distance from the boundary
a two-dimensional electron layer (or a double layer system).
We found that for
incompressible fractional quantum Hall states
the electric field $E$ of the test charge may deviate considerably from
unscreened field at distances
much  larger than the magnetic length.
The specific form of the dependence $E(r)$ is defined by the ground state of
the electron system.

\section{A monolayer system in an infinite medium}

To begin with, let us consider the problem of screening in an infinite
medium containing a two-dimensional electron layer in the fractional
quantum Hall regime. In order to describe the system, we consider the
model of spinless fermions
$\Psi$ (it is assumed that electrons are completely polarized in spin)
interacting with the two-dimensional Chern-Simons gauge field
$a_{\mu}$ and the
electromagnetic field $A_{\mu}$. The action of the system has the form
\begin{eqnarray}
S = S_{\rm CF}+S_{\rm em},
\label{1}
\end{eqnarray}
where
\begin{eqnarray}
S_{\rm CF} = \int d t d^2 r [
\Psi^*({\bf r}) (i \partial_t +\mu - a_0 - e A_0
\cr -
{1\over 2 m} (i \nabla_2 + {\bf a} + {e\over c} {\bf A}^{\rm pl})^2)
\Psi({\bf r})
+{1\over 2\pi \varphi } a_0 b]  \ ,
\label{2}
\end{eqnarray}
\begin{eqnarray}
S_{\rm em} = {1\over 8 \pi }\int d t d^3 r (\epsilon {\bf E}^2 - {\bf B}^2).
\label{3}
\end{eqnarray}
In Eq.~(\ref{2}), $m$ is the mass of composite fermions, $\mu$ the chemical
potential, $b = \partial_xa_y - \partial_ya_x$, the ``magnetic''
component of the gauge field,
$\varphi$, the number of gauge field flux quanta carried by a
composite quasiparticle ($\varphi$ is even). It is assumed that
the distribution of composite fermions along the $z$-axis is described
by  $delta$-function.
The transverse gauge is used for the field $a$ ($\partial_i a_i = 0$).
For the electromagnetic field also, we used the transverse gauge
in the plane ($\partial_x A_x + \partial_y A_y = 0$).

Functional integration with respect to the field $\Psi$ in the expression
for  the partition function of the system
\begin{eqnarray}
Z = \int D\Psi^* D\Psi D a_\mu \exp(i S)
\label{4}
\end{eqnarray}
gives the following effective action for a system of interacting gauge and
electromagnetic fields:
\begin{eqnarray}
S_{\rm eff}(a,A) = - i\ {\rm Tr}\ \log \ [
i \partial_t +\mu - a_0 - e A_0  \cr
-{1\over 2 m} (i \nabla_2 + {\bf a} + {e\over c} {\bf A}^{\rm pl})^2] +
\int d t d^2 r
{1\over 2\pi \varphi } a_0 b +S_{\rm em}.
\label{5}
\end{eqnarray}
The condition of stationary  configuration of the action~(\ref{5})
upon variation of the field
$a_{\mu}$ defines the value of the self-consistent effective field acting on
composite fermions:
\begin{eqnarray}
B_{\rm eff} =  B - {2 \pi c \varphi \over |e|} n_0,
\label{6}
\end{eqnarray}
where $n_0$ is the uniform electron concentration. The fractional quantum
Hall effect is observed for an integral number $N$ of filled Landau levels
in a field $B_{\rm eff}$, which corresponds to the filling factor
$\nu = N/(\varphi N +{\rm sign} B_{\rm eff})$.

We shall confine the subsequent analysis to a consideration of the part of
the effective action~(\ref{5}) that is quadratic in fluctuations of the
field $a_{\mu}$ and
$A_{\mu}$. In order to solve the problem considered here, we consider
only static fluctuations. Expansion of the action~(\ref{5})
in the vicinity of the stationary configuration gives
\begin{eqnarray}
S^{(2)}_{\rm eff}(a,A) = \cr ={1\over 2} \int d t d^2 q
[(a^*_{\mu {\bf q}}  + \tilde{A}^*_{\mu {\bf q}})
\Pi^\Psi_{\mu \nu q}
(a_{\nu {\bf q}}  + \tilde{A}_{\nu {\bf q}})\cr
+a^*_{\mu {\bf q}}
\Pi^{\rm CS}_{\mu \nu q}
a_{\nu {\bf q}}]  \cr
+{1\over 2} \int d t d^2 q  d q_z
A^*_{\mu {\bf q} q_z}
\Pi^{\rm em}_{\mu \nu q q_z}
A_{\nu {\bf q} q_z},
\label{7}
\end{eqnarray}
where the subscripts $\mu$ and $\nu$ assume the values 0 and 1 corresponding
respectively to zeroth and transverse components of the fields $a_{\mu}$ and
$A_{\mu}$ (in Eq.~(\ref{7}), we disregard the contribution of the component
$A_z$ which can be put equal to zero without any loss of generality in the
static problem under consideration), and ${\bf q}$ is the wave
vector component parallel to the $(x, y)$ plane. In Eq.~(\ref{7}),
$\tilde{A}_{0 {\bf q}} = e A_{0 {\bf q}}(z = 0)$,
$\tilde{A}_{1 {\bf q}} = (e/c) A_{1 {\bf q}}(z = 0)$,
\begin{eqnarray}
\Pi^\Psi_{\mu \nu q }  = -{1\over 2\pi \omega_{\rm c} }
\left( \matrix{q^2 \Sigma_0& i q \omega _{\rm c} \Sigma _1\cr
-i q \omega _{\rm c} \Sigma_1& \omega _{\rm c}^2(\Sigma _2+N)}\right),
\label{8}
\end{eqnarray}
\begin{eqnarray}
\Pi^{\rm CS}_{\mu \nu q} = {1\over 2\pi \varphi}
{\left( \matrix{0& i q\cr -i q&0} \right) } ,
\label{9}
\end{eqnarray}
\begin{eqnarray}
\Pi^{\rm em}_{\mu \nu q q_z}= {1\over 4\pi }
\left( \matrix{\epsilon (q^2+q^2_z)& 0\cr
0& - (q^2 + q^2_z)}\right).
\label{10}
\end{eqnarray}
In Eq.~(\ref{8}), we have introduced the notation
\begin{eqnarray}
\Sigma _j=-({\rm sign}B_{\rm eff})^j e^{-x}\sum_{n=0}^{N-1}
\sum_{m=N}^\infty
{n!\over m!}
{x^{m-n-1}\over { (m-n)}}  \cr
\times [L_n^{m-n}(x)]^{2-j} \cr \times
\left( (m-n-x)L_n^{m-n}(x) + 2x{d L_n^{m-n}(x)\over d x} \right) ^j   ,
\label{11}
\end{eqnarray}
where  $x=(ql_{\rm eff} )^2/2$, $l_{\rm eff} =(N/2\pi n_0)^{1/2}$
is the effective magnetic length, $\omega_{\rm c} = 2\pi
n_0/(m N)$ is the effective cyclotron frequency,
and $L_n^{m-n}(x)$ is the
generalized Laguerre polynomial. The quantities~(\ref{11}) are calculated
through the Green's current--current functions for the fermion system in the
field $B_{\rm eff}$
(the temperature is assumed to be equal to zero).
Expressions of the type~(\ref{11})
were first derived in the theory of anyons (see, for instance, \cite{9}).

Integration over fluctuations of the field $a$ leads to the following
expression for the action of the electromagnetic field:
\begin{eqnarray}
S(A) =  {1\over 4 \pi }
\int d t d q_z d q'_z d^2 q
A^*_{\mu {\bf q} q_z} \Lambda_{\mu \nu q}
A_{\nu {\bf q} q'_z }\cr  +{1\over 2}\int d t d q_z d^2 q
A^*_{\mu {\bf q} q_z} \Pi^{\rm em}_{\mu \nu q q_z }
A_{\nu {\bf q} q_z} ,
\label{12}
\end{eqnarray}
where
\begin{eqnarray}
\Lambda _{\mu \nu q} = -{e^2\over 2\pi \omega_{\rm c} \Delta _1}
\left( \matrix{q^2 \Sigma_0&
i {q \omega _{\rm c} D/c}\cr
-i {q \omega _{\rm c} D/c}&
{\omega _{\rm c}^2  (\Sigma _2+N)/c^2}}\right)
\label{13}
\end{eqnarray}
with
\begin{eqnarray}
D = \Sigma_1 +\varphi ( \Sigma_0(\Sigma_2+N) - \Sigma _1^2) ,
\label{14}
\end{eqnarray}
\begin{eqnarray}
\Delta_1 = (1-\varphi \Sigma _1)^2 -\varphi^2\Sigma_0(\Sigma_2+N) .
\label{15}
\end{eqnarray}

The action~(\ref{12}) leads to the following expression for the
electromagnetic field potential in a system with a test charge
$e_{\rm ext}$ placed at the origin:
\begin{eqnarray}
{1\over 2 \pi }
\int  d q'_z
\Lambda_{\mu \nu q}
A_{\nu {\bf q} q'_z } +
\Pi^{\rm em}_{\mu \nu q q_z }
A_{\nu {\bf q} q_z} =  \delta_{\mu 0} j_0 ,
\label{16}
\end{eqnarray}
where $j_0 = e_{\rm ext}/(2\pi)^{3/2}$.
The solution of Eq.~(\ref{16}) is sought in the
form
\begin{eqnarray}
A_{\mu {\bf q} q_z}={C_\mu (q)\over q^2 + q^2_z} .
\label{17}
\end{eqnarray}
Consequently, we obtain the following expression for the
quantity $A_{0 {\bf q} q_z}$:
\begin{eqnarray}
A_{0 {\bf q} q_z} =  {4 \pi j_0\over q^2+q^2_z}\cr
\times [ \epsilon  - {q e^4 c^{-2} \omega _{\rm c}
(\Sigma_0(\Sigma_2+N) - \Sigma _1^2)
+ q^2 e^2 \Sigma _0\over
e^2 c^{-2} \omega ^2_{\rm c} (\Sigma _2+N)+
q \omega _{\rm c} \Delta _1 }] ^{-1}
\label{18}
\end{eqnarray}

In the electrostatic limit $(c \to \infty)$,
Eq.~(\ref{18}) is reduced to the form
\begin{eqnarray}
A_{0 {\bf q} q_z} =  {4 \pi j_0 \over \epsilon}{1\over  q^2+q^2_z}
\left( 1 + {f_q \Sigma _0\over
\Delta _1-f_q \Sigma_0}\right),
\label{19}
\end{eqnarray}
where $f_q = e^2 q/ \epsilon \omega_{\rm c}$. Eq.~(\ref{19}) differs from
Eq.~(\ref{18}) significantly only for
\begin{eqnarray}
q < {e^2\omega _{\rm c}\over c^2} .
\label{20}
\end{eqnarray}
The values of $q$ satisfying this inequality are several orders of
magnitude lower than the characteristic scale of wave vectors of
the problem $\sim l_{\rm eff}^{-1}$. A consideration of the
difference between Eqs.~(\ref{18}) and (\ref{19}) in the region
(\ref{20}) leads to a very weak screening of the electric field of the test
charge at large distances. Here and below, we shall not analyze this very
weak effect, but confine to the approximate expression~(\ref{19}).
Note that this approximation corresponds to
negligible nondiagonal components of the tensor $\Lambda$ in Eq.~(\ref{16}),
which will be taken into consideration in the following sections.

The expression for the screened electric field of a test
charge, calculated from Eq. (\ref{19})
for $z = 0$, has the form
\begin{eqnarray}
E_{\rm pl}(r) = - {e_{\rm ext}\over \epsilon r^2 } (1+ F(r)),
\label{21}
\end{eqnarray}
where $r$ is the distance up to the test charge, and
\begin{eqnarray}
F(r)= r^2 \int dq J_1(q r){q f_q \Sigma_0\over
\Delta _1 -
f_q \Sigma_0}
\label{22}
\end{eqnarray}
($J_i(x)$ is the Bessel function).

To complete the picture, we also present the expression for the
$z$-component of the magnetic field (for $z = 0$), induced by the
test charge:
\begin{eqnarray}
\delta B_z(r) = - {e_{\rm ext} e^2\over c\epsilon }
\int d q q J_0(q r) {D\over
\Delta _1 -
f_q \Sigma_0}.
\label{23}
\end{eqnarray}
In this expression also, we have omitted the correction emerging in the
region~(\ref{20}). For characteristic values of the parameters and
for $e_{\rm ext} = e$, the numerical estimate for $\delta B_z$ is
found to be smaller than 1~Oe for all $r$, i.e., the effect is of
theoretical interest only. From the experimental point of view,
the most important effect is associated with the
deviation of $F(r)$ from zero in Eq.~(\ref{21}),
which may be of the order of unity
for finite values of $r$. The specific form of the
dependence $F(r)$ is determined  by the
ground state of the quantum Hall
system and is modified significantly upon a transition to another Hall step.
The following two sections are devoted to an analysis of this effect
in a semi-infinite medium.

\section{A monolayer system in a semi-infinite medium}

Let us now consider the geometry which seems to be appropriate for
observing the screening effect. We shall assume that the test charge
is located at the surface of a semi-infinite medium with the dielectric
constant
$\epsilon$. At a distance $a$ from the surface, the medium contains a
two-dimensional electron layer in the fractional quantum Hall
regime.We shall seek an expression for the screened field of a test
charge at the interface.

Disregarding the nondiagonal components of the tensor $\Lambda$, we can
present Eq.~(\ref{16}) in the geometry under consideration as follows:
\begin{eqnarray}
{1\over 2\pi }
\int  d q'_z (e^{i(q_z-q'_z)a}
\Lambda_{0 0 q}\cr + \epsilon _{q_z-q'_z} {q^2+q_z q'_z\over 2})
A_{0 {\bf q} q'_z}
=  j_0 ,
\label{24}
\end{eqnarray}
where
\begin{eqnarray}
\epsilon_{q_z} = {\epsilon +1\over 2} \delta (q_z)+ i
{\epsilon -1\over 2\pi } P({1\over q_z}).
\label{25}
\end{eqnarray}
The solution of Eq.~(\ref{24}) is sought in the form
\begin{eqnarray}
A_{0 {\bf q} q_z} = {C_1(q)+C_2(q) e^{i q_z a}\over q^2 + q^2_z} .
\label{26}
\end{eqnarray}
As a result, we arrive at the following expression for the
electric field projection on the plane $(x, y)$ at the interface:
\begin{eqnarray}
E_{\rm pl}(r) = - {e_{\rm ext}\over \epsilon' r^2 } (1+ F(r)) ,
\label{27}
\end{eqnarray}
where
\begin{eqnarray}
F(r)={\epsilon \over \epsilon'} r^2
\int dq J_1(q r){q f_q \Sigma_0 e^{-2 q a}\over
\Delta _1 -
f_q \Sigma_0 (1+{\epsilon -1\over \epsilon +1}e^{-2 q a})}
\label{28}
\end{eqnarray}
with $\epsilon' = (\epsilon + 1)/2$.

\begin{figure}
\narrowtext
\centerline{\epsfig{figure=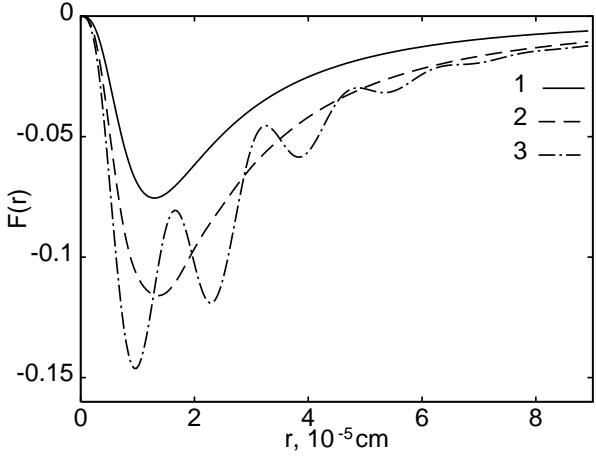,width=8cm}}
\vspace{0.5cm}
\caption{Relative screening of the test charge field by the
monolayer system.
1 - $\nu =1/3$; 2 - $\nu =3/7$; 3 - $\nu =5/11$.}
\label{fig1}
\end{figure}

The dependence $F(r)$ is shown in Fig.\ref{fig1} for
the filling factors $\nu$ = 1/3, 3/7, 5/11. Calculations were made by
using the following values of the parameters:
$n_0 = 10^{11}$ cm$^{-2}$,
$m=0.25 m_e$, $\epsilon =12,6$,
$a=500\AA$.
It can be seen from Fig \ref{fig1} that the system
has a significant screening of the electric field of the test charge
at distances considerably larger than the magnetic length.
As the filling factor $\nu$ approaches the value 1/2, the
dependence $F(r)$ becomes oscillatory. Note that an increase
in the value of $a$ weakens the effect. A decrease in this parameter
leads not only to an increase in the amplitude of the effect, but also
to a wider range of filling factors for which $F(r)$ oscillates as a
function of the distance. In particular, numerical computations for the
geometry considered in Sec.~I lead to an oscillatory dependence $F(r)$
for all filling factors corresponding to a fractional
quantum Hall effect. In order to observe these oscillations, one can
measure the electric field inside the dielectric medium. For the case
considered in this section, oscillations emerge when the effective
magnetic length becomes of the order of $2a$.

\section{A double-layer system in a semi-infinite medium}

Let us now consider the screening of the field of a test charge by a
double-layer electron system. We shall consider a system in which the
states described by the Halperin's wave function\cite{5} are realized.
Such a system can be described by introducing two types of Chern--Simons
fields corresponding to statistical charges belonging to composite
quasiparticles in opposite layers, and to an additional term in the
Lagrangian of the system, which is nondiagonal in the gauge
fields. The action of the system has the form
\begin{eqnarray}
S_{\rm CF} = \int d t d^2 r \sum_{k=1}^2 [
\Psi^*_k({\bf r}) (i \partial_t +\mu - a_{k0} - e A_{k0}
\cr - {1\over 2 m} (i \nabla_2  + {\bf a}_k +
{e\over c} {\bf A}^{\rm pl}_k)^2) \Psi_k({\bf r})\cr
+\sum_{k,k'=1}^2 \Theta_{k k'} a_{k0} b_{k'}],
\label{29}
\end{eqnarray}
where $A_{k 0}$ and ${\bf A}_k^{\rm pl}$ are the scalar and
vector components of the electromagnetic field potential in the
layer $k$, and
\begin{eqnarray}
\Theta  _{k k' }=
{1\over 2\pi (\varphi^2 -s^2)}
{\left( \matrix{\varphi &- s\cr -s&\varphi } \right) } .
\label{30}
\end{eqnarray}
In this equation, $\varphi$ and $s$ are the numbers of the gauge field
flux quanta carried by a composite quasiparticle, which correspond to the
statistical charges of quasiparticles in the same layer and in the
opposite layer respectively ($\varphi$ is even while $s$ is an arbitrary
integer). For the sake of simplicity, we
confine the analysis to two equivalent layers. Fractional quantum
Hall effect in the system~(\ref{29}) is realized for filling
factors  $\nu =2 N/((\varphi +s)N +{\rm sign}B_{\rm eff})$
($\nu=2\nu_i$, where $\nu_i$
is the filling factor per layer). Carrying out a procedure analogous
to the one carried out  in Sec.~I, we arrive at the following
expression for the action of the electromagnetic field of the
system~(\ref{29}):
\begin{eqnarray}
S(A) = {1\over 2} \int d t d^2 q  A^*_{k \mu {\bf q}}
\Lambda_{k k' \mu \nu  q} A_{k' \nu {\bf q}} + S_{\rm em},
\label{31}
\end{eqnarray}
where
\begin{eqnarray}
\hat{\Lambda}_{k k'} ={1\over 2}\left( \matrix{
\hat{\Lambda}^+ +\hat{\Lambda}^-&\hat{\Lambda}^+ -\hat{\Lambda}^-\cr
\hat{\Lambda}^+ - \hat{\Lambda}^-&\hat{\Lambda}^+ +\hat{\Lambda}^-
}\right) .
\label{32}
\end{eqnarray}
Matrices $\hat{\Lambda}^+$ and $\hat{\Lambda}^-$ in Eq.~(\ref{32}) are
defined by Eqs.~(\ref{13})--(\ref{15}) in which the parameter
$\varphi$ is replaced by $\varphi + s$ and $\varphi - s$ respectively.
Note that although Eqs.~(\ref{29}) and (\ref{30})
become meaningless for $\varphi = s$, the expressions~(\ref{31})
and (\ref{32}) remain valid even in this case.
The situation $\varphi = s$ corresponds to infinite rigidity
of the antiphase oscillations of the gauge fields. Going over in
Eqs. (\ref{29}) and (\ref{30}) to new variables corresponding to
synphase and antiphase oscillations, we must consider
for the integration variables in the case $\varphi = s$ only
synphase oscillations of the fields $a_{i\mu}$ and put the
variable corresponding to antiphase oscillations
equal to zero. It can be verified that in this case also we arrive at the
relations (\ref{31}) and (\ref{32}).

For the geometry in which the $z$-coordinate of the first and second layer
is equal to $-a$ and $-(a + d)$ respectively, we obtain from
Eq. (\ref{31}) the following equation for the scalar potential of the
system:
\begin{eqnarray}
{1\over 4 \pi }\int  d q'_z \{e^{i(q_z-q'_z)a}
[(\Lambda^+_{0 0 q} + \Lambda^-_{0 0 q})
(1+e^{i(q_z-q'_z)d}) \cr +
(\Lambda^+_{0 0 q} - \Lambda^-_{0 0q})
(e^{-i q'_z d}+e^{i q_z d})] \cr
+ \epsilon _{q_z-q'_z} ({q^2+q_z q'_z})\}
A_{0 {\bf q} q'_z}
=  j_0 ,
\label{33}
\end{eqnarray}
where the nondiagonal components of the tensors $\Lambda^+$ and $\Lambda^-$
are neglected as before.

The solution of Eq.~(\ref{33}) is sought in the form
\begin{eqnarray}
A_{0 {\bf q} q_z} = {C_1(q)+C_2(q) e^{i q_z a}+C_3(q) e^{i q_z (a+d)}
\over q^2 + q^2_z} .
\label{34}
\end{eqnarray}
As a result, we arrive at an expression for
the electric field projection on the plane $(x, y)$
for $z =0$, whose form coincides with Eq.~(\ref{27}) in which the
function $F(r)$ is modified to the form
\begin{eqnarray}
F(r)={\epsilon \over \epsilon'} r^2
\int dq J_1(q r){q S_q e^{-2 q a}\over
R_q -
{\epsilon -1\over \epsilon +1} S_q e^{-2 q a})} ,
\label{35}
\end{eqnarray}
where
\begin{eqnarray}
R_q = (\Delta_1^+  - f_q E^+_q \Sigma_0)
(\Delta_1^-  - f_q E^-_q \Sigma_0),
\label{36}
\end{eqnarray}
\begin{eqnarray}
S_q = f_q \Sigma _0 [{1\over 2} (E^+_q)^2 \Delta_1^-+
{1\over 2} (E^-_q)^2 \Delta_1^+ \cr -  f_q E^+_q E^-_q \Sigma_0] .
\label{37}
\end{eqnarray}
In formulas (\ref{36}) and (\ref{37}), the functions
$\Delta_1^+$ and $\Delta_1^-$ are
defined by formula~(\ref{13}) in which $\varphi$ is replaced
by $\varphi + s$ and
$\varphi - s$ respectively, and $E_q^{\pm} = (1 \pm e^{-qd})$.

It can be seen from Eqs.~(\ref{35})--(\ref{37}) that screening in a
two-layer system depends on parameter $\varphi + s$ as well as $\varphi - s$.
Hence for a fixed value of the filling factor which depends only on the
first of these parameters, the expression~(\ref{35}) differs for the
cases $s = 0$ and $s \neq 0$. Consequently, a transition from a state
with $s = 0$ to a state with nonzero $s$ (such a state corresponds to
the Halperin's wave function) will be manifested in a variation of the
dependence of screened field of a test charge on the distance.

\begin{figure}
\narrowtext
\centerline{\epsfig{figure=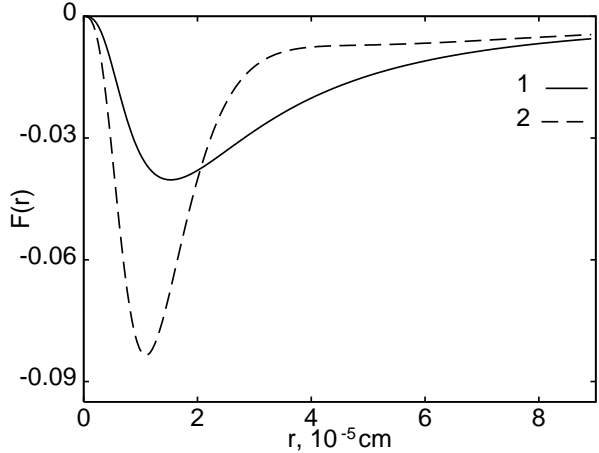,width=8cm}}
\vspace{0.5cm}
\caption{Relative screening of the test charge field by the
double-layer system for $\nu = 2/5$. 1 - $\varphi = 4$, $s = 0$;
2 - $\varphi = 2$, $s = 2$.}
\label{fig2}
\end{figure}

Let us consider a system with $\nu = 2/5$. In such a system,
the fractional quantum Hall effect may correspond to sets of parameters
($\varphi = 4$ and $s = 0$) and ($\varphi = 2$ and $s = 2$).
Fig.~\ref{fig2} shows the dependence $F(r)$ for these two cases
for the same values of the system parameters as in Sec.~II and for
$d = 400\,{\rm \AA}$. It can be seen from the curves that the dependences
$F(r)$ differ considerably for these two cases. An experimental
observation of such a sharp variation in screening upon a slight variation
in the separation between layers points towards a phase transition
between different ground states in a two-layer system. Note that the
situation considered in this work differs from the one considered
by us in Ref.~\cite{10} where we studied screening in a two-layer system
in an infinite medium with two test charges located in the opposite
layers. In such a case, the screening of the test charge field depends
only on parameter $\varphi + s$, and the variation of the spatial
distribution of the induced charge  during a phase transition between
generalized Laughlin states is associated just with a variation
of this parameter with a simultaneous reversal of the sign of
$B_{\rm eff}$. Only a few of the possible transitions satisfy this
condition. In particular, the transition considered above for $\nu = 2/5$
(which is most suitable for observation since it corresponds to lowest
level in the hierarchy of the generalized Laughlin states)
does not satisfy such a condition. The origin of the effect
considered in this work is associated with the asymmetric arrangement
of the test charge relative to the two-layer system. In particular,
this is manifested in that a decrease in the separation between layers
leads to a suppression of the effect.

Thus, we have considered in this work the screening of the electric field of
a test charge by a monolayer and a two-layer systems of composite fermions.
The expressions for the screened field are obtained by
taking into account the effect of the interface. It is shown that a
partial screening of the test charge electric field occurs in the system
at distances much larger than the magnetic length. The specific form of the
dependence of the screened field on the distance from the test charge
is modified considerably upon a variation of the ground state of
the system. The observation of the screening effect as a function
of the filling factor and separation between layers (in a two-layer system)
can be treated as a possible experimental verification
of the model of composite fermions and the method of observing changes
in the topological order in fractional quantum Hall systems. The solutions
obtained in this work pertain to the case when the test charge and
the electric field detector are placed on the surface of the sample
containing a two-dimensional electron  layer. The approach used in
this work can be modified to describe a different geometry of the
experiment.

\end{multicols}
\end{document}